\documentclass[english]{article}
\usepackage[T1]{fontenc}
\usepackage[latin9]{luainputenc}
\usepackage{geometry}
\usepackage{amstext}
\usepackage{amssymb}
\usepackage{feyn}
\usepackage{tikz}
\usepackage{graphicx}
\usepackage{array}
\usepackage{booktabs}

\usepackage{amsmath}
\usepackage{amsfonts}

\usepackage[normalem]{ulem}
\usepackage{color}
\usepackage{listings,braket}
\usepackage{caption}
\usepackage{subcaption}
\usepackage{float}
\definecolor{darkgreen}{rgb}{0,0.35,0}
\definecolor{Rood}{rgb}{1, 0, 0}

\makeatletter
\usepackage[english]{babel}
\usepackage{feyn}

\makeatother

\begin{document}

\title{\textbf{ { A study of the  violation of the Bell-CHSH inequality through a pairing mechanism}}}

{\author{\textbf{S.~P.~Sorella$^1$}\thanks{silvio.sorella@gmail.com}\\\\\
\textit{{\small $^1$UERJ -- Universidade do Estado do Rio de Janeiro,}}\\
\textit{{\small Instituto de F\'{\i}sica -- Departamento de F\'{\i}sica Te\'orica -- Rua S\~ao Francisco Xavier 524,}}\\
\textit{{\small 20550-013, Maracan\~a, Rio de Janeiro, Brasil}}\\
}

\date{}

\maketitle
\begin{abstract}
The violation of the Bell-CHSH inequality for bipartite systems is discussed by making use of the pseudospin operators which enable us to group all modes of the Hilbert space of the system in pairs. We point out that a single pair can be already employed to perform a test of the Bell-CHSH inequality in order to check out its violation. The mechanism is illustrated with the help of $N00N$ states as well as with coherent and squeezed states.

\end{abstract}

\section{Introduction}\label{intro}
The Bell-CHSH inequality \cite{Bell:1964kc,Clauser:1969ny} is a fundamental tool of  Quantum Mechanics. Its violation signals the existence of strong  correlations which cannot be accounted for by local realistic theories. It means a drastic departure from any possible pre-deterministic type of description. {\color{red} Bell-CHSH's inequality plays a ivotal role in many areas. To quote a few of them we may refer to: quantum computation, quantum information, quntum cryptography as well as to the more recent study of the entanglement in systems exhibiting topological phases \cite{cil1,cil2}.} \\\\There are two main ingredients in the elaboration of the Bell-CHSH inequality, which we write in its conventional form as 
\begin{equation}
 \langle \psi | {\cal C}_{CHSH} | \psi \rangle  =  \langle 
 \psi  (A_1+A_2)B_1 + (A_1-A_2)B_2| \psi \rangle     \;,  \label{BCHSH}
\end{equation}
One speaks of a violation of the Bell-CHSH inequality whenever 
\begin{equation} 
|  \langle \psi | {\cal C}_{CHSH} | \psi \rangle | > 2 \;. \label{vb}
\end{equation}
As it is apparent from eq.\eqref{BCHSH}, the first ingredient for a possible violation is a judicious choice of the  state $|\psi \rangle$ characterizing the bipartite system. The second important and more difficult task is the explicit choice of the four dichotomic bounded Hermitian operators $(A_i, B_k)$, $i,k=1,2$, which, according to \cite{tsi1}, have to fulfill the strong requirements
\begin{equation} 
A^2_i=1 \;, \qquad B^2_k =1 \;, \qquad [A_i, B_k] = 0 \;. \label{req}
\end{equation}
To our knowledge, although the choice of the state $|\psi\rangle $ relies essentially on its entanglement features, the issue of the choice of the operators $(A_i, B_k)$ is still a matter of investigation, see  \cite{Gisin,Peruzzo:2023nrr} for  recent accounts.  \\\\In the sequel, we shall provide a simple construction of $(A_i, B_k)$ based on a pairing mechanism relying on the so-called pseudospin operators, which enable us to introduce dichotomic  Bell operators in Hilbert spaces of infinite dimension. \\\\More precisely, following \cite{psi1,psi2,psi3}, one introduces the operators 
\begin{equation} 
s_x = \sum_{n=0}^\infty s^{(n)}_x \;, \qquad s_y = \sum_{n=0}^\infty s^{(n)}_y \;, \qquad s_z = \sum_{n=0}^\infty s^{(n)}_z \label{spin1} 
\end{equation}
where 
\begin{eqnarray} 
s^{(n)}_x & = & \vert 2n+1 \rangle \langle 2n \vert + \vert 2n \rangle \langle 2n+1 \vert \;, \nonumber \\ 
s^{(n)}_y & = &i\left(  \vert 2n+1 \rangle \langle 2n \vert - \vert 2n \rangle \langle 2n+1 \vert \right) \;, \nonumber \\ 
s^{(n)}_z & = & \vert 2n+1 \rangle \langle 2n+1 \vert - \vert 2n \rangle \langle 2n \vert. \label{spin2}
\end{eqnarray} 
with $\{ \vert n \rangle, n=0,1....,\infty\}$ being the usual number basis. 
A simple  calculation shows that 
\begin{equation} 
\left[s^{(n)}_x,s^{(n)}_y \right] = 2 i s^{(n)}_z \;, \qquad
\left[s^{(n)}_y,s^{(n)}_z \right] = 2 i s^{(n)}_x \;, \qquad 
\left[s^{(n)}_z,s^{(n)}_x \right] = 2 i s^{(n)}_y.  \label{spin4}
\end{equation} 
Therefore, it follows that these operators obey the same algebraic relations of the spin $1/2$ Pauli matrices: 
\begin{equation} 
\left[ s_x,s_y \right] = 2 i s_z \;, \qquad
\left[s_y,s_z \right] = 2 i s_x \;,\qquad
\left[s_z,s_x \right] = 2 i s_y.  \label{spin5}
\end{equation} 
from which the name {\it pseudospin} follows. \\\\From expressions \eqref{spin2} one observes that the introduction of the pseudospin operators can be related to a pairing mechanism in the Hilbert space, a pair being given by two modes, namely,  $(\vert2n \rangle, \vert2n+1\rangle)$. Each pair of modes gives rise to a set of operators, $(s^{(n)}_x,s^{(n)}_y,s^{(n)}_z)$, which obey the spin $1/2$ algebra.  \\\\The observation of the pairing mechanism  goes back to the {\color{red} seminal work of \cite{Gisina,Gisin}. More precisely, in \cite{Gisina} it has been shown that any entangled two-qbits, {\it i.e.} any pair of modes, yields a violation of the Bell-CHSH inequality. In the light of Gisin's results \cite{Gisina} the novelty of the present work relies on a quite simple and systematic construction of the Bell operators $(A_i,B_k)$, see eqs.\eqref{BellOptwomodes}, enabling us to establish a direct bridge with the aforementioned pseudospin operators. As such, the present construction provides  concise and helpful applications to Hilbert spaces of infinite dimensions, as discussed in Sections \eqref{Coherent} and \eqref{ho}. Recently, the pairing mechanism  has been employed in \cite{Peruzzo:2023nrr},  where it has been shown that, in the case of two spin $j$ particles,  each single pair might already be used in order to  test the Bell-CHSH inequality. }\\\\
In the sequel, we shall analyze the Bell-CHSH inequality by considering a single pair of modes, showing that it already leads to a violation of the Bell-CHSH inequality. This  provides  a  quick and useful framework  in order to check out the existence of possible violations. \\\\Picking up only one mode means that the Bell operators act non-trivially only on a single pair identified,  for example, by the modes $(\vert 0 \rangle, \vert 1 \rangle)$. Let $\vert x, y \rangle$ stand for a generic basis element of the Hilbert space ${\cal H}_a \otimes {\cal H}_b$  of the bipartite system. For the Bell operators $(A,B)$ we shall set 
\begin{align}\label{BellOptwomodes}
	A \vert 0, y \rangle &= e^{i \alpha}  \vert 1, y \rangle; \,\, 	A \vert 1, y \rangle = e^{-i \alpha} \vert 0, y \rangle; \, \forall y, \nonumber \\
	B \vert x, 0 \rangle &= e^{i \beta}  \vert x, 1\rangle; \, 	B \vert x, 1 \rangle = e^{-i \beta} \vert x, 0 \rangle; \, \forall x, 
\end{align}
and acting as the identity on all the other states, {\it i.e.},
\begin{align}\label{BellOptwomodes2}
	A \vert x, y \rangle &= \vert x, y \rangle, \quad \forall x \geq 2, \nonumber \\
	B \vert x, y \rangle &= \vert x, y \rangle, \quad \forall y \geq 2.
\end{align}
The quantities $(\alpha, \beta)$ are arbitrary parameters, to be chosen at the best convenience. These parameters play a role akin to that of the four angles of the spin 1/2 Bell-CHSH inequality.  One sees that the operator $A$ acts only on the first entry of $\vert x, y \rangle$, while the operator $B$ only on the second one. In terms of the pseudospin operators, it turns out that the operator $A$ can be written as 
\begin{equation} 
A = \left( {\vec u} \cdot {\vec s}^{(0)} + {\cal R} \right) \otimes I 
\end{equation}
where $\vec{u}$ denotes the unit vector 
\begin{equation}
{\vec u} =\left( \cos(\alpha), \sin(\alpha),0 \right) \;, \qquad {\vec u} \cdot {\vec u} = 1, \label{vecu}
\end{equation} 
and ${\cal R}$ is the identity operator for $x\ge 2$:
\begin{equation} 
{\cal R} = \sum_{n=2}^\infty \vert n \rangle \langle n \vert.    \label{Rop}
\end{equation}
Analogous expressions can be  written down for $B$, $A'$ and $B'$. For the primed operators, the parameters $\alpha$ and $\beta$ are simply replaced by $\alpha'$ and $\beta'$. It is immediate to check that the required properties, eq.\eqref{req},  for the Bell-type operators are satisfied. \\\\As we shall see in the following examples, the one pair setup described by equations \eqref{BellOptwomodes}, \eqref{BellOptwomodes2} yields an efficient way to find out the violation of the Bell-CHSH inequality. \\\\The work is organized as follows. In Sect.\eqref{setup} we illustrate the setup by analyzing the example of a generic bipartite system. In Sect.\eqref{N00N-state} we discuss the $N00N$ states. Sections\eqref{Coherent},\eqref{ho} deal with coherent and squeezed states, respectively. Section \eqref{conc} collects our conclusion.

\section{Illustrating the  setup} \label{setup} 
As already underlined in the Introduction, the framework can be described by the following steps: 
\begin{itemize} 
\item Let ${\cal H}_a$ be a $d_a$-dimensional Hilbert space. The dimension $d_a$ can be either finite or infinite. We mirror now the starting Hilbert space and introduce a second Hilbert space ${\cal H}_b$ with the same dimension of ${\cal H}_a$, {\it i.e.} $d_a=d_b$. The resulting Hilbert space is ${\cal H} = {\cal H}_a \otimes {\cal H}_b$. \\\\As a helpful example, we might take an Harmonic oscillator of frequency $\omega_a$ and add a second oscillator with frequency $\omega_b$, so as to have two distinguishable systems, while keeping the same dimensionality for the corresponding Hilbert spaces.  Similarly, one can consider  two spin $j$ particles with different masses, and so on. 
\item Let us denote by $\{| n_a\rangle\}$ and $\{|m_b\rangle\}$  orthonormal basis  in ${\cal H}_a$ and ${\cal H}_b$, respectively. \\\\Pick up  two different elements of $\{|n_a\rangle\}$, say $(\eta_a, \chi_a)$:
\begin{equation} 
\langle \eta_a | \eta_a \rangle=1 \;, \qquad \langle \chi_a | \chi_a \rangle=1 \;, \qquad \langle \eta_a | \chi_a \rangle=0 \;. \label{etachi}
\end{equation} 
\item Introduce the pure state $| \psi \rangle $ defined as 
\begin{equation} 
| \psi \rangle = \frac{1}{\sqrt{2}} \left( |\eta_a\rangle |\chi_b\rangle + |\chi_a\rangle |\eta_b\rangle \right) \;. \label{psi}
\end{equation} 
The above expression has a very transparent meaning: we have taken one single mode of the basis $\{| n_a\rangle\}$ and we have entangled it with a single mode of the basis  $\{|m_b\rangle\}$. 
\item Define  now the four dichotomic operators $(A_i, B_k)$ as 
\begin{eqnarray} 
A_i | \eta_a \rangle = e^{i \alpha_i} | \chi_a \rangle \;, \qquad A_i | \chi_a \rangle = e^{-i \alpha_i} | \eta_a \rangle \;, \qquad A_i = 1\;\;  {\it  on \; the\; remaining \; elements \;of \{| n_a\rangle\} } \;, \nonumber \\[2mm]
B_k | \eta_b \rangle = e^{i \beta_k} | \chi_b \rangle \;, \qquad B_k | \chi_b \rangle = e^{-i \beta_k} | \eta_b \rangle \;, \qquad B_k = 1\;\;  {\it  on \; the\; remaining \; elements \;of \{| m_b\rangle\} } \;, \label{ABop}
\end{eqnarray} 
where $(\alpha_1, \beta_k)$ stand for arbitrary real parameters. The operators $(A_i, B_k)$ defined in that way are Hermitian and fulfill the requirements \eqref{req}. 
\item An elementary calculation shows that 
\begin{equation} 
\langle \psi | A_i B_k |\psi \rangle = \cos(\alpha_i-\beta_k) \;. \label{ABpsi}
\end{equation}
\end{itemize} 
Therefore, for the Bell-CHSH inequality, one gets 
\begin{equation} 
\langle \psi | {\cal C}_{CHSH} | \psi \rangle = \left( \cos(\alpha_1-\beta_1) + \cos(\alpha_2-\beta_1) + \cos(\alpha_1-\beta_2) - \cos(\alpha_2-\beta_2) \right)  \;,\label{res}
\end{equation} 
Choosing 
\begin{equation} 
\alpha_1 = 0 \;, \qquad \alpha_2 = \frac{\pi}{2} \;, \qquad \beta_1= \frac{\pi}{4} \;, \qquad \beta_2= - \frac{\pi}{4} \;, \label{choice} 
\end{equation} 
one finds maximum violation, namely: the saturation of Tsirelson's bound  \cite{tsi1} 
\begin{equation}
\langle \psi | {\cal C}_{CHSH} | \psi \rangle = 2 \sqrt{2} \;. \label{mv}
\end{equation} 
We see thus that, entangling one single mode of ${\cal H}_a$ with one single mode of ${\cal H}_b$, results in the maximum violation of the Bell-CHSH inequality, due to the choice of the operators $(A_i, B_k)$. 

\section{The $N00N$ state} \label{N00N-state}
As first example, let us discuss the number state, called the $N00N$ state, see \cite{Gerry} and refs. therein: 
\begin{equation} 
| \psi \rangle = \frac{1}{\sqrt{2}} \left( | N_a\rangle | 0_b\rangle + |0_a\rangle |N_b\rangle \right) \;, \label{N00N}
\end{equation} 
where 
\begin{eqnarray} 
|N_a\rangle & = & \frac{1}{\sqrt{N!}} (a^\dagger)^N |0_a\rangle \;, \qquad |N_b\rangle = \frac{1}{\sqrt{N!}} (b^\dagger)^N |0_b\rangle \;, \nonumber \\
\left[a, a^{\dagger}\right] & =&  \left[b, b^{\dagger}\right] =1 \;, \qquad \left[a, b\right]=0 \;. \label{cab}
\end{eqnarray} 
Following the previous construction, for the operators $(A_i, B_k)$, we write 
\begin{eqnarray} 
A_i | N_a \rangle = e^{i \alpha_i} | 0_a \rangle \;, \qquad A_i | 0_a \rangle = e^{-i \alpha_i} | N_a \rangle \;, \qquad A_i = 1\;\;  {\it  on \; the\; remaining \; elements \;of \;the \;basis} \;, \nonumber \\[2mm]
B_k | N_b \rangle = e^{i \beta_k} | 0_b \rangle \;, \qquad B_k | 0_b \rangle = e^{-i \beta_k} | N_b \rangle \;, \qquad B_k = 1\;\;  {\it  on \; the\; remaining \; elements \;of \;the \;basis } \;, \label{ABopN00N}
\end{eqnarray}
Therefore, the choice \eqref{choice}  leads to the maximum violations of the Bell-CHSH inequality, {\it i.e.}
\begin{equation}
\langle \psi | {\cal C}_{CHSH} | \psi \rangle = 2 \sqrt{2} \;. \label{mvN00N}
\end{equation}

\section{Coherent states}\label{Coherent}

It is worth observing that the procedure previously outlined can be  generalized to the more complex case in which $|\eta\rangle$ and $|\chi\rangle$  are not orthogonal, {\it i.e.} $\langle \eta | \chi \rangle \neq 0$. In this case, it suffices to expand $|\eta\rangle$ and $|\chi\rangle$ along the orthonormal basis and pick up two specific modes. \\\\Let us illustrate the procedure by considering, for instance, the state \cite{Gerry}
\begin{equation} 
| \psi \rangle  = \frac{1}{\sqrt{2}} \frac{1}{\sqrt{1+|\alpha|^2 e^{-|\alpha|^2}}}\left( | 1_a\rangle | \alpha_b\rangle + |\alpha_a\rangle |1_b\rangle \right) \;,\label{coh}
\end{equation} 
where $| \alpha \rangle$ denotes a coherent state, {\it i.e.}
\begin{equation}
| \alpha \rangle = e^{-\frac{|\alpha|^2}{2}}\; e^{\alpha a^{\dagger}} |0\rangle = e^{-\frac{|\alpha|^2}{2}}\; \sum_{n=0}^{\infty} \frac{\alpha^n}{\sqrt{n!}} | n\rangle \;. \label{coherent}
\end{equation}
Besides the mode $|1\rangle$, we might pick up now the mode $|0\rangle$.  We rewrite  thus expression 
\eqref{coh} as 
\begin{equation} 
| \psi \rangle  = \frac{ e^{-\frac{|\alpha|^2}{2}}}{\sqrt{2}}  \frac{1}{\sqrt{1+|\alpha|^2 e^{-|\alpha|^2}}}\left(  |1_a\rangle |0_b\rangle +2 \alpha |1_a\rangle |1_b\rangle  +  |0_a\rangle |1_b\rangle  
+ |1_a\rangle \sum_{n=2}^{\infty} \frac{\alpha^n}{\sqrt{n!}} | n_b\rangle 
+ |1_b\rangle \sum_{n=2}^{\infty} \frac{\alpha^n}{\sqrt{n!}} | n_a\rangle \right) \;,\label{coh1}
\end{equation} 
For the operators  $(A_i, B_k)$, we have 
\begin{eqnarray} 
A_i | 0_a \rangle = e^{i \alpha_i} | 1_a \rangle \;, \qquad A_i | 1_a \rangle = e^{-i \alpha_i} | 0_a \rangle \;, \qquad A_i = 1\;\;  {\it  on \; the\; remaining \; elements \;of \;the \;basis} \;, \nonumber \\[2mm]
B_k | 0_b \rangle = e^{i \beta_k} | 1_b \rangle \;, \qquad B_k | 1_b \rangle = e^{-i \beta_k} | 0_b \rangle \;, \qquad B_k = 1\;\;  {\it  on \; the\; remaining \; elements \;of \;the \;basis } \;. \label{ABopcoh}
\end{eqnarray}
Therefore 
\begin{eqnarray} 
A_i B_k | \psi \rangle & = & \frac{ e^{-\frac{|\alpha|^2}{2}}}{\sqrt{2}}  \frac{1}{\sqrt{1+|\alpha|^2 e^{-|\alpha|^2}}}\left(
e^{-i( \alpha_i -\beta_k)} |0_a\rangle |1_b\rangle 
+  e^{i( \alpha_i -\beta_k)}  |1_a\rangle |0_b\rangle  +
 2 \alpha e^{-i( \alpha_i +\beta_k)} |0_a\rangle |0_b\rangle  \right) \nonumber \\
 & + & \frac{ e^{-\frac{|\alpha|^2}{2}}}{\sqrt{2}}  \frac{1}{\sqrt{1+|\alpha|^2 e^{-|\alpha|^2}}}\left(
 e^{-i \alpha_i} |0_a\rangle \sum_{n=2}^{\infty} \frac{\alpha^n}{\sqrt{n!}} | n_b\rangle 
+ e^{-i \beta_k} |0_b\rangle \sum_{n=2}^{\infty} \frac{\alpha^n}{\sqrt{n!}} | n_a\rangle \right) \;,\label{coh2}
\end{eqnarray} 
so that, for the Bell-CHSH inequality one finds 
\begin{equation}
\langle \psi | {\cal C}_{CHSH} | \psi \rangle =   e^{-|\alpha|^2} \frac{1}{(1+|\alpha|^2 e^{-|\alpha|^2})}\left( \cos(\alpha_1-\beta_1) + \cos(\alpha_2-\beta_1) + \cos(\alpha_1-\beta_2) - \cos(\alpha_2-\beta_2) \right)  \;. \label{cohBCHSH} 
\end{equation} 
Setting 
\begin{equation} 
\alpha_1 = 0 \;, \qquad \alpha_2 = \frac{\pi}{2} \;, \qquad \beta_1= \frac{\pi}{4} \;, \qquad \beta_2= - \frac{\pi}{4} \;, \label{choicecoh} 
\end{equation} 
 one gets 
 \begin{equation}
\langle \psi | {\cal C}_{CHSH} | \psi \rangle = e^{-|\alpha|^2} \frac{1}{(1+|\alpha|^2 e^{-|\alpha|^2})}2 \sqrt{2} \;. \label{cohBCHSHcohfn} 
\end{equation} 
 which exhibits a violation whenever 
 \begin{equation} 
  \frac{ e^{-|\alpha|^2}}{(1+|\alpha|^2 e^{-|\alpha|^2})}> \frac{1}{\sqrt{2}} \;. \label{small}
 \end{equation} 
 Condition \eqref{small}  results in a rather small  violation, due to the presence of the damping factor   $e^{-|\alpha|^2}$.  This is in nice agreement with the experimental findings of \cite{Gerry},  being understood in terms of the known features of the coherent states,  often employed in the study of the semiclassical limit of Quantum Mechanics. Their entanglement properties are known to be delicate and, somehow,  fragile,  see \cite{Sanders} for a review.
 
 \section{Squeezed states}\label{ho} 
As last example, let us consider the case of the squeezed state:  
\begin{equation} 
|\eta \rangle = \sqrt{(1-\eta^2)} \;e^{\eta a^\dagger b^\dagger} |0 \rangle \;, \qquad \langle \eta | \eta \rangle = 1 \;,  \label{etast}
\end{equation} 
where the real parameter $\eta$ is constrained to belong to the interval 
\begin{equation} 
0 < \eta < 1 \;. \label{intv} 
\end{equation} 
The operators $(a,b)$ obey the following commutation relations: 
\begin{eqnarray} 
 \left[ a, a^{\dagger}\right] & = & 1\;, \qquad [a^{\dagger}, a^{\dagger}] =0 \;, \qquad  [a, a] =0 \;, \nonumber \\
 \left[ b, b^{\dagger}\right]  & = & 1\;, \qquad [b^{\dagger}, b^{\dagger}] =0 \;, \qquad  [b, b] =0 \;, \nonumber \\
 \left[a, b \right] & = &  0\;, \qquad [a, b^{\dagger}] =0 \;,  \label{ccrqm}
\end{eqnarray} 
with
\begin{equation} 
a |0\rangle = b |0\rangle = 0 \;. \label{st}
\end{equation}
In order to study the violation of the Bell-CHSH inequality, we proceed as before and pick up the modes $|0 \rangle$ and $|1 \rangle$ in expression \eqref{etast}, namely, we write 
\begin{equation} 
|\eta \rangle = \sqrt{(1-\eta^2)} \left( |0_a\rangle |0_b \rangle  + \eta  |1_a\rangle |1_b \rangle  + \sum_{n=2} \eta^n |n_a\rangle |n_b \rangle \right) \;,  \label{etast1}, 
\end{equation} 
where 
\begin{equation} 
|n_a\rangle |n_b \rangle = \frac{1}{n!} (a^\dagger)^n (b^\dagger)^n |0\rangle \;, \qquad |0\rangle = |0_a\rangle|0_b\rangle \;. \label{vc}
\end{equation} 
Similarly to the previous case, for the operators  $(A_i, B_k)$, we have 
\begin{eqnarray} 
A_i | 0_a \rangle = e^{i \alpha_i} | 1_a \rangle \;, \qquad A_i | 1_a \rangle = e^{-i \alpha_i} | 0_a \rangle \;, \qquad A_i = 1\;\;  {\it  on \; the\; remaining \; elements \;of \;the \;basis} \;, \nonumber \\[2mm]
B_k | 0_b \rangle = e^{i \beta_k} | 1_b \rangle \;, \qquad B_k | 1_b \rangle = e^{-i \beta_k} | 0_b \rangle \;, \qquad B_k = 1\;\;  {\it  on \; the\; remaining \; elements \;of \;the \;basis } \;. \label{ABopcoh}
\end{eqnarray}
{\color{red} 
Therefore, 
\begin{equation} 
A_i B_k | \eta \rangle = \sqrt{(1-\eta^2)} \left( e^{i(\alpha_i +\beta_k)} | 1_a \rangle| 1_b \rangle + \eta e^{-i(\alpha_i +\beta_k)} | 0_a \rangle| 0_b \rangle +  \sum_{n=2} \eta^n |n_a\rangle |n_b \rangle \right) \;. \label{add1}
\end{equation}
From 
\begin{equation} 
\sum_{n=2} \eta^{2n} = -1 - \eta^2 + \frac{1}{1-\eta^2} \;, \label{add2} 
\end{equation} 
one gets 
}
\begin{equation} 
\langle \eta | A_i B_k | \eta \rangle = 1 + (1-\eta^2) \left(  2 \eta \cos(\alpha_i + \beta_k)  -1 -\eta^2 \right)  \;. \label{etaAB}
\end{equation} 
Setting 
\begin{equation} 
\alpha_1 = 0 \;, \qquad \alpha_2 = \frac{\pi}{2} \;, \qquad \beta_1= -\frac{\pi}{4} \;, \qquad \beta_2=  \frac{\pi}{4} \;, \label{choicesq} 
\end{equation} 
 for the Bell-CHSH inequality one gets 
 \begin{equation}
\langle \eta | {\cal C}_{CHSH} | \eta  \rangle = 2 + 2 (1-\eta^2) \left( 2 \sqrt{2}\; \eta -1 -\eta^2 \right)\;. \label{cohBCHSHsq} 
\end{equation}
 There is violation whenever 
 \begin{equation} 
 \sqrt{2} - 1 < \eta < 1 \;. \label{vsq}
 \end{equation} 
 The maximum value of the violation occurs for $\eta \approx 0.7$, yielding 
 \begin{equation}
\langle \eta | {\cal C}_{CHSH} | \eta  \rangle \approx 2.5\;. \label{cohBCHSHsqv} 
\end{equation} 

\section{Conclusion}\label{conc}
In this work, we have discussed  a  setup for the study of the  violation of the Bell-CHSH inequality in bipartite systems described by pure states, namely
\begin{equation} 
| \psi \rangle = \frac{1}{\sqrt{2}} \left( |\eta_a\rangle |\chi_b\rangle + |\chi_a\rangle |\eta_b\rangle \right) \;. \label{psiconc}
\end{equation}
We have pointed out that, besides the choice of the state itself, $|\psi\rangle$, the Bell-CHSH inequality depends on the way one introduces the operators $(A_i, B_k)$. Although in the original spin 1/2 example treated by Bell, the operators $(A_i,B_k)$ turn out to be uniquely defined in terms of Pauli matrices, their introduction is much more involved in  the case of other higher dimensional Hilbert spaces. \\\\In our construction, we have distinguished two cases:
\begin{itemize}
\item  $|\eta_a \rangle$ and $ |\chi_a\rangle$ are orthogonal, {\it i.e.} $\langle \eta_a | \chi_a \rangle=0$. In this case, the construction of the four operators $(A_i,B_k)$ devised in equation \eqref{ABop} leads immediately to the maximum violation, {\it i.e} to Tsirelson's bound $2\sqrt{2}$. 
\item when $|\eta_a \rangle$ and $ |\chi_a\rangle$ are not orthogonal, {\it i.e.} $\langle \eta_a | \chi_a \rangle \neq 0$, one expands them in the respective basis and one picks up two specific orthogonal modes. The construction of the  operators $(A_i,B_k)$ of  eq.\eqref{ABop} enables us  to investigate in a rather quick and simple ways the violation of the Bell-CHSH also in this more complex case, as illustrated in the examples of the coherent and squeezed states. 
\end{itemize}
{\color{red} As last remark let us mention that, in principle, given a density matrix $\rho$  representing a mixed state, the present setup may be employed to discuss the Bell-CHSH inequality as well. Let us give a simple example of this by considering the Werner state for two spin $1/2$, namely 
\begin{equation}
\rho = \frac{1-w}{4} 1_a \otimes 1_b + w |\psi\rangle \langle \psi |  \;, \label{werner} 
\end{equation}
where $|\psi \rangle$ stands for the Bell singlet 
\begin{equation} 
|\psi\langle = \frac{1}{\sqrt{2}}\left( |+\rangle_a \otimes |-\rangle_b - |-\rangle_a \otimes |+\rangle_b \right) \;, \label{werner2}
\end{equation}
and $w$ is the Werner parameter, $0<w<1$. Use of the equations \eqref{BellOptwomodes2} yields immediately 
\begin{equation} 
{\rm Tr}({\cal C}_{CHSH} \;\rho) = w 2\sqrt{2} \;, \label{werner3} 
\end{equation}
implying violation of the Bell-CHSH inequality when $w > 1/\sqrt{2}$. This result is in agreement with the standard treatment of the mixed states via the positive partial transpose (PPT) criterium. Though, it remains the nontrivial challenge of establishing a clear connection between the definition of the Bell operators given in \eqref{BellOptwomodes2} and the entanglement features of generic mixed states for higher dimensional Hilbert spaces, a task beyond the aim of the present work. 
}

\section*{Acknowledgements}
The authors would like to thank the Brazilian agencies CNPq and FAPERJ for financial support.  S.P.~Sorella is a level $1$ CNPq researcher under the contract 301030/2019-7.

\end{document}